# Broadband Perfect Absorption of Ultrathin Conductive Films with Coherent Illumination: Super Performance of Electromagnetic Absorption

Sucheng Li[1], Jie Luo[1], Shahzad Anwar[1], Shuo Li[1], Weixin Lu[1], Zhi Hong Hang[1], Yun Lai[1], Bo Hou[1], Mingrong Shen[1], and Chinhua Wang[1,2]

*[1] College of Physics, Optoelectronics and Energy,*
*Soochow University, 1 Shizi Street, Suzhou 215006, China*
*[2] Jiangsu Key Lab of Advanced Optical Manufacturing Technologies & Collaborative Innovation Center of Suzhou Nano Science and Technology,*
*Soochow University, 1 Shizi Street, Suzhou 215006, China*

Absorption of microwave by metallic conductors is exclusively inefficient, though being natively broadband, due to the huge impedance mismatch between metal and free space. Reducing the thickness to ultrathin conductive film may improve the absorbing efficiency, but is still bounded by a maximal 50% limit induced by the field continuity. Here, we show that broadband perfect (100%) absorption of microwave can be realized on a single layer of ultrathin conductive film when it is illuminated coherently by two oppositely incident beams. Such an effect of breaking the 50% limit maintains the intrinsic broadband feature from the free carrier dissipation, and is frequency-independent in an ultrawide spectrum, ranging typically from kilohertz to gigahertz and exhibiting an unprecedented bandwidth close to 200%. In particular, it occurs on extremely subwavelength scales, ~λ/10000 or even thinner, which is the film thickness. Our work proposes a way to achieve total electromagnetic wave absorption in a broadband spectrum of radio waves and microwaves with a simple conductive film.

## 1. Introduction

The dissipation of microwaves in conductive materials, such as metal and semiconductor, is caused by the free carriers colliding with the ion lattice, and could be broadband in nature, due to the absence of the bounding force for the free carriers [1]. Thus, a natively broadband microwave absorber could be designed by straightforwardly utilizing the unpatterned conductive materials in first thought. However, the good conductors themselves, for example noble metals, reflect nearly all incoming microwaves with negligible absorptions. From the impedance point of view, this absorption inefficiency is due to the huge impedance mismatch between the metal and the free space blocks the microwave from penetrating the metal in the first place. By reducing the thickness, researchers have found the ultrathin conductive film of conductivity $\sigma$ and thickness $h$ may dissipate significantly electromagnetic (EM) waves in a frequency-irrelevant manner because of the improved match between the film impedance (that is, sheet resistance $R_s=1/\sigma h$ in the ultrathin cases, measuring the resistance along the plane of the film of arbitrarily sized square shape, also expressed in *ohms per square*) and the vacuum impedance $Z_0$ [2-6]. However, the calculation shows the absorption is bound within 50% whatever the sheet resistance is, and it is hence assumed that the 50% limit can not be broken under the ultrathin approximation ($h<<\delta$, the latter being skin depth and hence the parallel component of electric field being continuous across the film). The limit severely compromises the appealing properties of inherently broadband and ultrathin characteristics when using the conductive film as absorbers.

Since a broadband ultrathin perfect absorber is highly desirable in numerous applications ranging from microwave to infrared and visible light [7-10], various strategies of designing ultrathin perfect absorbers are proposed, representatively including the thick substrate or lossy element addition [3,9-10], the composite structure/material employment [2,11-14], the local resonance introduction [15-20], and the functional reflection boundary usage [21-24] . Consequently, the absorption can be enhanced up to 100%, but the resultant absorber either tends to become wavelength bulky if its performance is broadband, or might require complicated fabrications and usually works only over the narrow band or the single frequency if its thickness is subwavelength. To qualify the absorber's performance, the relative bandwidth of the perfect absorption is defined as $2(f_u-f_l)/(f_u+f_l)$, where $f_u$ and $f_l$ are the upper and lower frequencies of the 100% absorption band, respectively, and the definition gives the limiting bandwidth 200%. The typical relative bandwidths among the existed perfect absorber designs are usually less than 100% and never approach the limiting value 200% [7-24].

In recent years, a new concept called coherent perfect absorption (CPA) which can accomplish 100% perfect absorption has attracted a lot of research interest [25-34]. Being regarded as the time-reversal process of laser generation, CPA takes place when a Fabry-Perot (FP) dielectric cavity is illuminated by two counter-propagating light beams satisfying coherence conditions, and all input EM energy is trapped and dissipated inside the cavity, resulting in 100% absorption. Originally based on the FP cavity configuration, the optical CPA operates to a specific frequency and the wavelength thick materials. Very recently, the metallic film CPA has been investigated theoretically, exhibiting the unparalleled bandwidth and low profile advantages [30].

In another important technological field, the conductive thin films are being used very popularly and broadly nowadays, especially integrated as transparent conductors with optical transparency for various optoelectronic and custom electronic devices [35-43]. Furthermore, there have been enormous developments in the mobile internet with wireless EM signals in radio frequency (RF) and microwave domain. Therefore, the absorption of conductive films or transparent conductors in RF/microwave surroundings has been a topic of recent interest. In this paper, we experimentally demonstrate that the broadband perfect absorption of relative bandwidth 100% can be realized on the ultrathin conductive films illuminated coherently by two opposite microwave beams. The film samples are the standard transparent conductors with the conductive layer thickness, i.e. the absorbing thickness, being ~$\lambda$/10000 ($\lambda$ denoting wavelength in vacuum). The coherent illumination brings the ultrathin film absorbance up to 100% in a frequency-independent manner, which implies an ultra-broadband behavior in the frequency spectrum of radio waves and microwaves. In principle, the lower frequency, $f_l$, of the perfect absorption band can be close to zero hertz and the relative bandwidth can be nearly 200%. This extremely subwavelength scaled, inherently broadband, perfect absorption which excludes any extra addition or engineering of complicated structures or thick substrates not only breaks the 50% dissipation limit, but also represents two significant advantages in EM absorption research: the probably simplest in absorber design and the most efficient in performance.

## 2. Broadband Perfect Absorption

It has been found that the ultrathin conductive films have the maximal 50%, frequency-independent absorption to the incoming EM wave [2-6]. The optimal absorption depends solely on the sheet resistance, requiring $R_s=Z_0/2=188\Omega$ at normal incidence, and is not related to the specific material composition of the films. In this work, we use the customized commercial transparent conductive films with the 2.6um thick conducting layer to demonstrate the broadband perfect absorption phenomenon (see Supplementary Material A). As illustrated by the inset in Fig. 1(a), the transparent conductive films have the 0.2mm thick plastic substrate. Their sheet resistance was characterized by the four-probe instrument. In order to obtain the reflection and transmission spectra, two identical double-ridged horn antennas are connected with two ports of the network analyzer to generate and receive microwave in free space (see Supplementary Material B). The sample is inserted into the microwave passage for measuring S11 and S21 parameters, i.e. reflection/transmission coefficients, $r$ and $t$. Then the absorbance is calculated via the relation $A=1-R-T$, where the reflectance $R=|r|^2$ and the transmittance $T=|t|^2$. The measured results at normal incidence are plotted in Fig. 1(a), which displays the absorption is in its maximum, consistent with its sheet resistance.

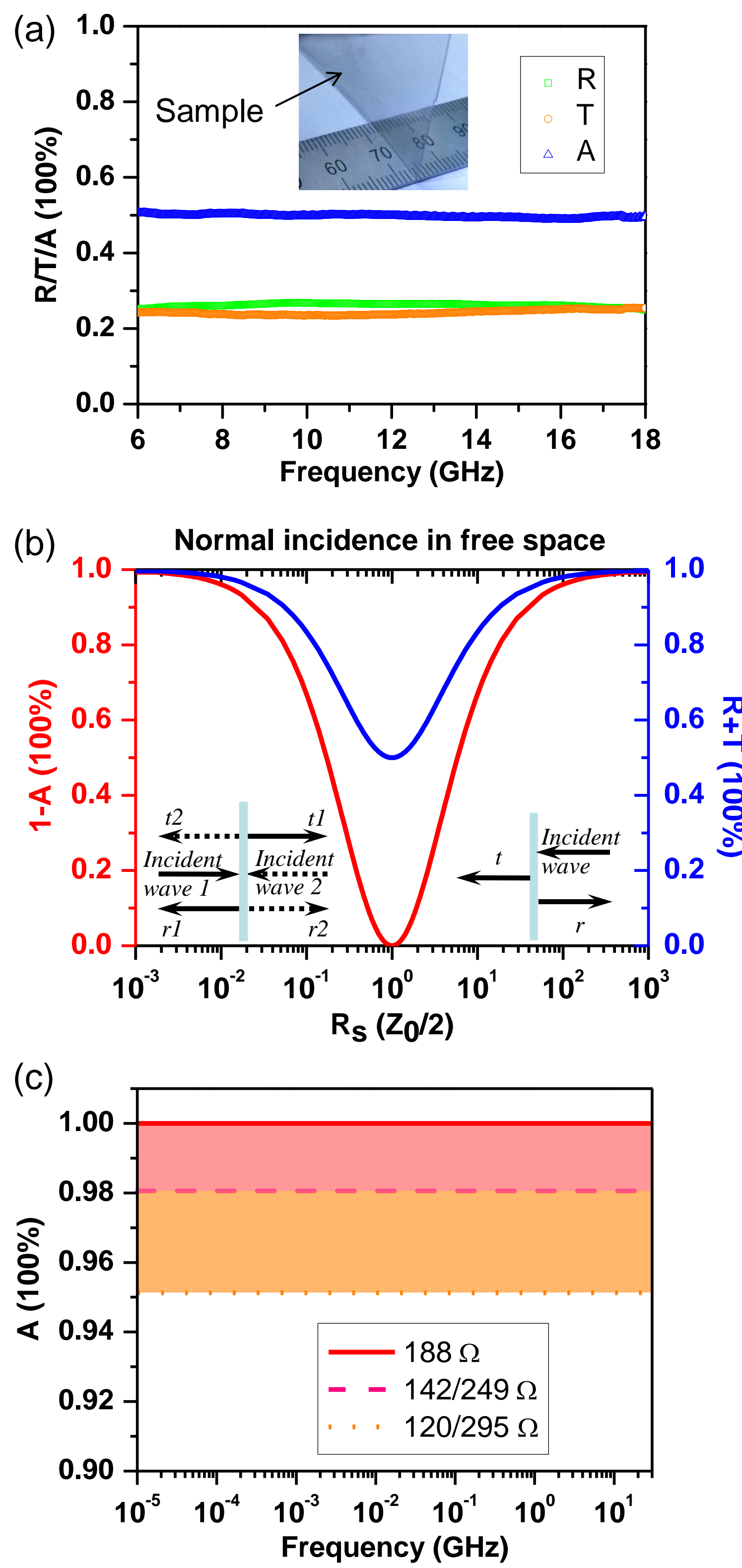

**FIGURE 1| The 50% absorption limit and the frequency-independent perfect absorption.** (a) The microwave reflectance, transmittance, and absorbance of the

ultrathin conductive film. Its sheet resistance is 180Ω. The inset is the photo of the sample, where the underlying ruler has the minimum division, *millimeter*. (b) The effect of the sheet resistance on the reflectance and transmittance in two cases, the single beam illumination (right inset, blue curve and right axis) and the coherent dual beams illumination (left inset, red curve, and left axis). (c) The frequency-independency of the broadband absorption. Three absorbance curves are calculated according to Eq. (1), denoted by the sheet resistance. All the shading area represents $A \geq 95\%$, if $120\Omega \leq R_s \leq 295\Omega$; the red area represents $A \geq 98\%$, if $142\Omega \leq R_s \leq 249\Omega$; the frequency-independent perfect absorbance $A=100\%$, if $R_s=188\Omega$.

When the optimal absorption takes place, the transmission and reflection coefficients are $t1=0.5$ and $r1=-0.5$, respectively. In this situation, we launch the second beam which is incident from the right side and has the same amplitude and phase as the first beam, as depicted schematically by the left inset in Fig. 1(b). Then, the second beam also has the transmission $t2=0.5$, completely cancelling with the reflection $r1=-0.5$ of the first beam. Likewise, the perfect cancellation happens at the right side for the transmission of the first beam and the reflection of the second beam. Thus, the total incident energy is entirely deposited on the film and dissipated by the free carriers, which is just the coherent perfect absorption.

From the calculation (see Supplementary Material C), the reflectance in the coherent illumination case is:

$$R = |1-(0.5Z_0/R_s)|^2 / |1+(0.5Z_0/R_s)|^2 . \quad (1)$$

For the coherent illumination, all scattered signals, including $r1$, $t1$, $r2$, and $t2$, are considered as reflection, and the absorbance is calculated to be $1-R$. Figure 1(b) shows the calculated quantity $1-A$ with varying the sheet resistance in the dual beams and the single beam case at normal incidence. It is seen that $R=0$ ($A=1$) at $R_s=Z_0/2$ for the coherent dual beams illumination, whereas the absorbance is only 50% under the single beam case. In particular, both Eq. (1) and the CPA condition, $R_s=1/\sigma h=Z_0/2$, shows no physical quantities depending on frequency, if $\sigma$ is non-dispersive, leading to the frequency independent absorption behavior which is robust in the radio/microwave regime where the DC conductivity is applicable and might deteriorate in terahertz domain where $\sigma$ becomes dispersive. Therefore, the effect in Fig. 1(b) is not specific to some frequency, but general in a broadband spectrum of radio waves and microwaves. In Fig. 1(c), we calculate the absorbance curves for several $R_s$ values and explicitly illustrate the frequency-independency of the intrinsically broadband absorption over the giant frequency range from kHz to GHz. Note that >98% absorbance can be obtained while $R_s$ ranges extensively from 142Ω to 249Ω, which indicates the resistance insensitivity of the microwave perfect absorption.

The basic difference between the optical CPA and our case is the material system. In the former, a high-Q resonant cavity made of the low-loss semiconductors is necessary for the anti-lasering process, and thus leads inevitably to the selectivity of working frequency from the cavity size and the sensitivity to material parameters in terms of dispersive permittivity or refractive index. In our case, no cavity resonance is involved, and the lossy conductive film is non-resonant and therefore endows the frequency-independency which is very appealing to the ultra-broadband applications. Furthermore, in our discussed EM spectrum range where both ultrathin condition and low frequency condition hold, the dimensional parameter and the material parameter of the film can be combined into a single nondispersive characteristic quantity, $R_s$, which has the explicit meaning in physics and the great utility in engineering applications (see the use of the impedance language in Supplementary Material C). In addition, in a reverse process of the microwave CPA, the system does not behave like any maser [44], but a non-resonant sheet antenna that radiates the plane wave, once driven by external sources.

## 3. Experiments in free space and waveguide

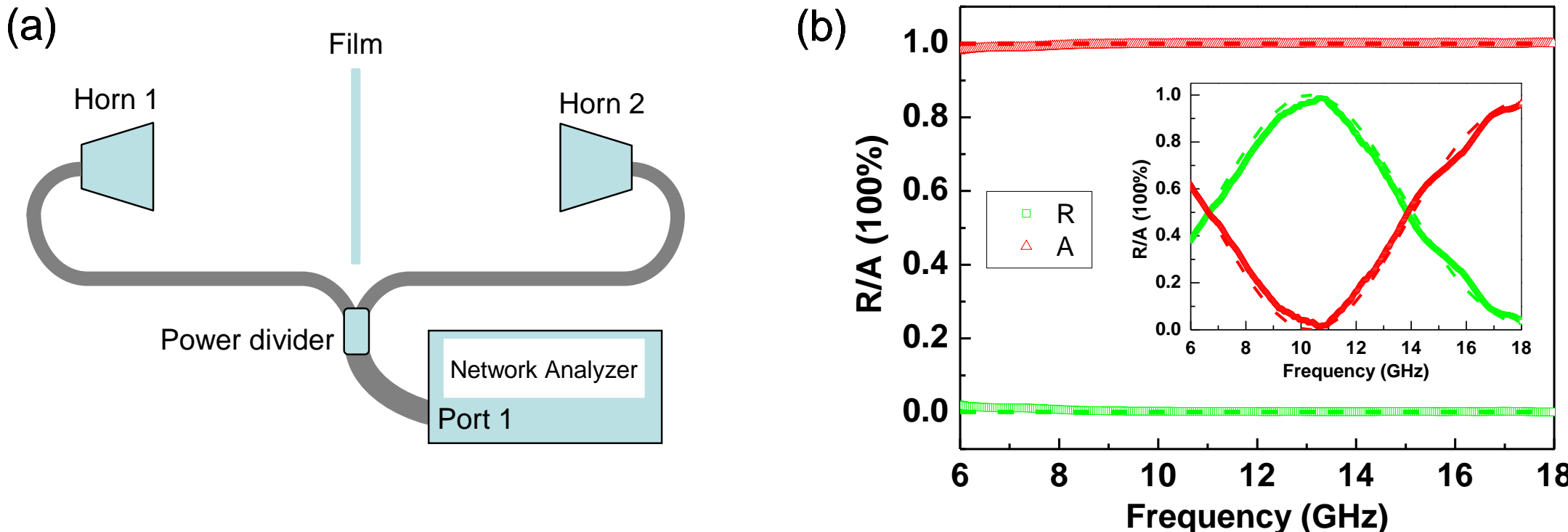

**FIGURE 2| The coherent perfect absorption in free space.** (a) The schematic drawing of the experimental setup in free space. (b) The measured (open symbols) and calculated (dash lines) results in free space for the cases of $L$=0 and $L$=14.5mm (inset). $L$ denotes the difference in the distance of the film to the two horn antennas. The sample has the sheet resistance 180Ω.

The CPA experiment is implemented first in free space, and the measuring setup is schematically drawn in Fig. 2(a). Two identical rectangular horn antennas are connected to a -3dB power divider (the broadband Wilkinson type [45]) via two coaxial cables of the same length. Then the power divider is connected with port 1 of the network analyzer. The sample is first inserted in the proper position with the same distance to the antennas, i.e. the distance difference $L$=0. In this symmetrical configuration, the two incident waves emitted from the horns have the same amplitude and phase when they reach the sample, satisfying the coherent illuminating condition for CPA. The measured and calculated results are plotted in Fig. 2(b), where the absorbance (reflectance) curve is noticed to be nearly 100% (0%) over the experimental frequency band from 6 to 18GHz. This complete absorption is seen to be frequency-independent, corresponding to a relative bandwidth 100%. Its experimental bandwidth may approach 200%, if not limited by our measuring capacity. As discussed above, the native broadband feature comes from the nondispersive conductivity and is an intrinsic consequence of free carriers dissipating the EM wave of low frequencies. A detailed analyze of the EM parameters of the experimental film sample is given in Supplementary Material D. In particular, note that the effective dissipating layer is only ~3um thick, which is on extremely subwavelength scale

(~λ/10000).

If the distances of the sample to the two antennas are changed so that *L* is nonzero, say 14.5mm, the experimental spectra are shown in the inset of Fig. 2(b). It is seen that a ~0 absorption minimum appears. This zero absorption is related to the relative phase, Δφ, of two incident waves when they hit the sample. If $\Delta\varphi=k_0L=m\pi$ ($k_0=2\pi f/c$, *f* being frequency, *c* being the speed of light in vacuum, *m* being odd number) at the specific frequency, the waves cancel each other at the sample position, no free carrier dissipation is induced, and consequently there is no absorption at all. The absorption minimum position, 10.5GHz, is in good agreement with the condition $k_0L=\pi$.

The CPA can also take place in the off-normal incidence. In Fig. 3(a), the two beams are illuminating the sample at the same oblique angle $\theta_0$. Like the free space case, owing to the mutual cancellation of the reflected and transmitted beams on either side, the 100% EM energy carried by the dual oblique beams is trapped on the conductive film and finally dissipated via ohmic loss. The reflectance under both TM and TE incidences are calculated and plotted in Figs. 3(b) and 3(c), where the green lines denote the zero reflectance conditions, $R_s=(Z_0/2)\cos\theta_0$ for TM illumination or $R_s=(Z_0/2)/\cos\theta_0$ for TE illumination. These two conditions are seen not to depend on frequency. With the ~$Z_0/2$ sheet resistance, the angular tolerance for the >90% TM&TE absorbance is seen as large as 60 degree. Instead of doing the oblique CPA experiments in free space with four horn antennas, we perform the measurement inside the rectangular waveguide, and the resultant difference is merely that $\theta_0$ changes with frequency according to the relation $\cos\theta_0=\beta/k_0$ where $\beta$ is the propagation constant of the waveguiding mode.

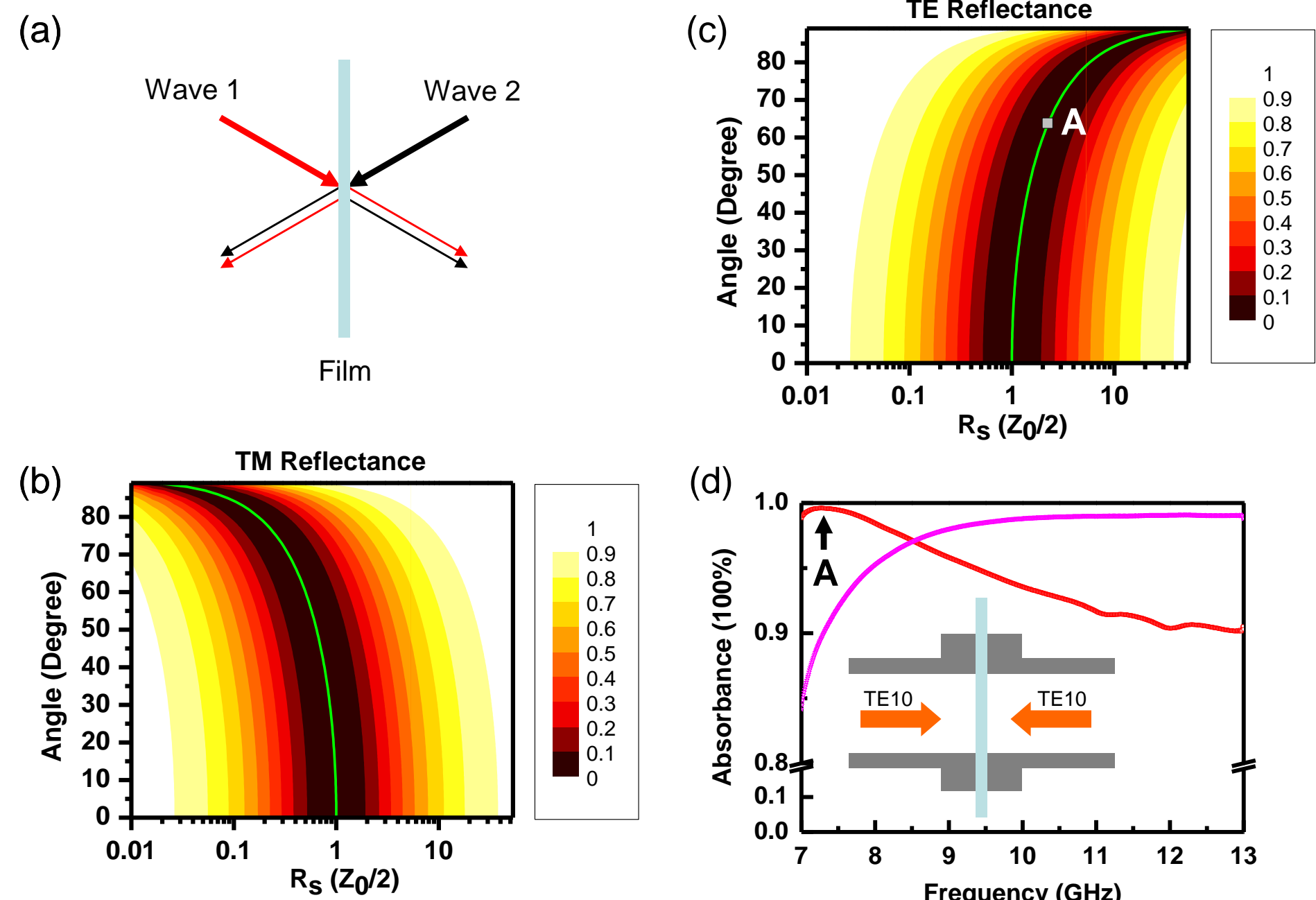

**FIGURE 3| The coherent perfect absorption in oblique incidence.** (a) The schematic drawing of the CPA at oblique incidence of angle $\theta_0$. (b) and (c) The calculated reflectance under TM and TE illumination, respectively, where the green line denotes the CPA condition, $R_s = (Z_0/2)\cos\theta_0$ (TM) and $R_s = (Z_0/2)/\cos\theta_0$ (TE). (d) The measured results in the X-band waveguide, as illustrated schematically by the inset. The samples have the sheet resistance 221Ω (magenta line) and 442Ω (red line).

In the waveguide experiment, the two identical coax-to-waveguide adapters (X-band WR90) are connected with port 1 of the network analyzer via the power divider. The sample is sandwiched between the two adapters, and the TE10 mode is launched inside the $22.86\times10.16\text{mm}^2$ rectangular waveguide from both sides and travels toward the sample with the same phase, as illustrated by the inset in Fig. 3(d). To make sure the single-mode operation, the measuring frequencies are below 13GHz which is determined by the TE20 cutoff. The magenta line is the result of the sample $R_s$=221Ω, and indicates the absorption increases from 84% at 7GHz to 99% beyond 10GHz. In alternative view, the TE10 mode can be decomposed as two TEM beams propagating at the same oblique angle $\theta_0=\cos^{-1}(\beta/k_0)$. The TE10 mode will be asymptotic to the normally incident TEM wave with increasing frequency, which means $\theta_0$ decreases with frequency. Therefore, provided that the sheet resistance is close to $Z_0/2$, lower frequencies will see ~15% reflectance, and higher frequencies will see less reflectance. The magenta line is seen approaching 100% towards higher frequencies, which is consistent with Fig. 3(c).

The sample of $R_s$ =442Ω is also tested in our waveguide setup, and the result is plotted as the red line in Fig. 3(d). It is noted that the absorbance over the X-band is greater than 90% and the nearly full absorption, 99.6%, appears around 7.35GHz, denoted as the label "A" in Fig. 3(d). As shown in Fig. 3(c), the perfect absorption for $R_s$ =442Ω happens to the 64 degree oblique angle, seeing the label, which corresponds to 7.3GHz TE10 mode. The calculated results of the two samples agree very well with the measured ones and are not plotted here. Therefore, the waveguide experiment proves that the broadband perfect absorption phenomenon under coherent illumination persists in the off-normal incidence, even at the considerable oblique angle.

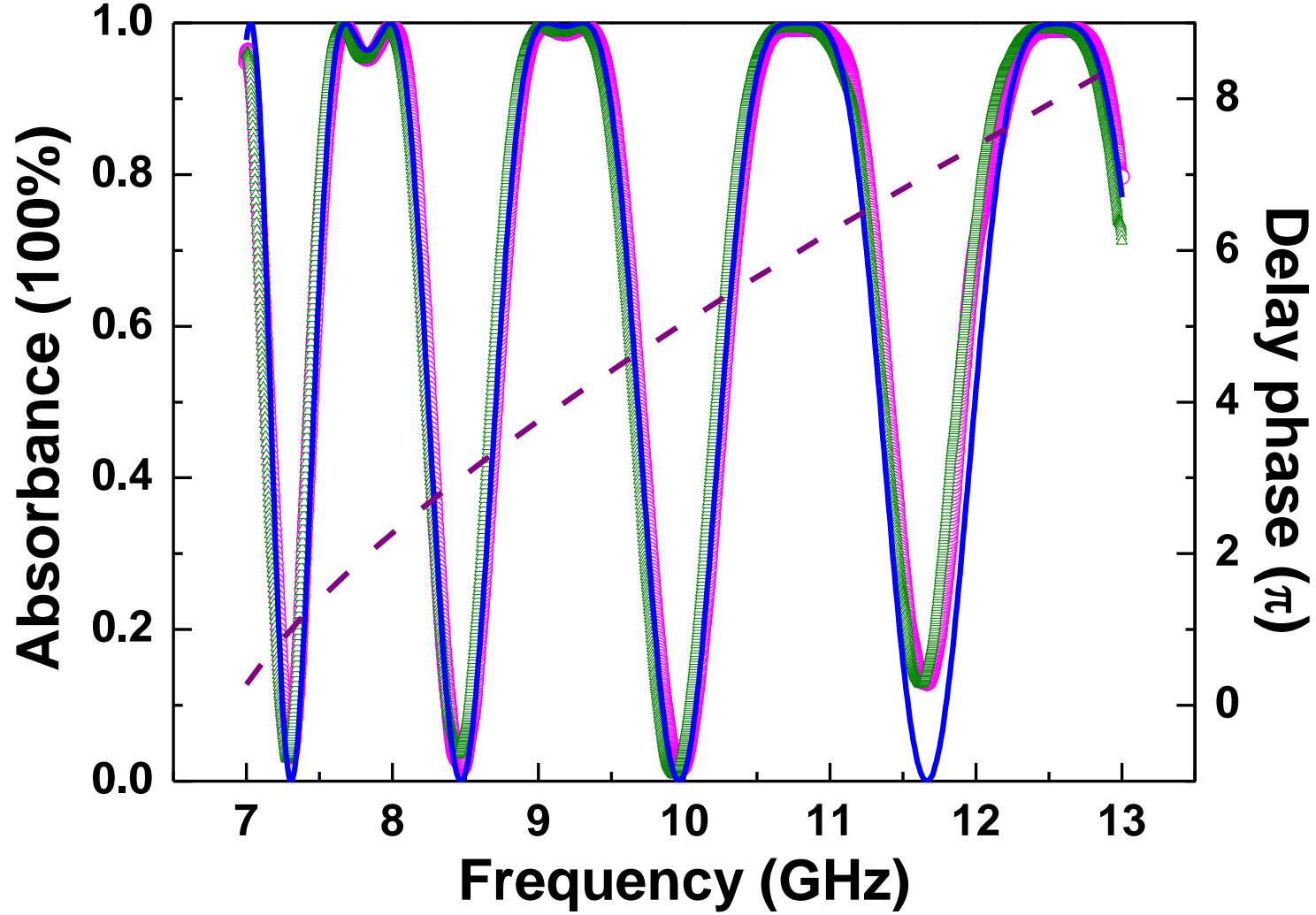

**FIGURE 4| The coherent perfect absorption in the waveguide with large phase delay.** The measured (open symbols) and calculated (solid line) absorbance in the waveguide with the $L$=140mm long delay line addition. The dash line denotes the delay phase $\beta L$, and is associated with the right axis. The sample has the sheet resistance 221Ω.

The relative phase of the two incident beams when reaching the sample can be controlled via adding the fixed delay line to the passage of one beam. This can be done with a standard section of X-band waveguide (length $L$=140mm) in our waveguide experiments. (The phase delay introduced by this waveguide is $\beta L$. For the general tunability purpose, a phase shifter may be used to provide arbitrary phase control.) First, we inserted it between the left waveguide adapter and the sample, and measured the absorbance, shown as green symbols in Fig. 4. Then we inserted it between the right waveguide adapter and the sample, and measured the spectra, shown as magenta symbols in Fig. 4. As expected, both curves agree very well, and reveal the rapid modulation of absorption from 0 to 1 within the X-band. The absorbance was calculated and plotted as solid line in Fig. 4. Interestingly, the absorption twin-peak appears and evolves into the plateau with frequency, giving rise to the relatively wide 100% absorbing bands. These bands as well as the absorption minima come from the relative phase condition $\Delta\varphi=\beta L=m\pi$ (the even number $m$ for the band and the odd $m$ for minima), as indicated by the dash line in Fig. 4. The calculation (see Supplementary Material C) reveals that the condition fulfilling the absorption minima is $\cos(\beta L)=-1$. The condition associated with the absorption bands is $(Z_0/2)\cos\Delta\varphi/(R_s\cos\theta_0)=1$, holding at two frequencies around $\Delta\varphi=m\pi$ ($m$ being the even number), and a detailed discussion about the two frequencies, i.e. the twin absorption peaks, is given in Supplementary Material E. The absorption modulation and the band formation in the waveguide system will find applications in designing new microwave devices like waveguide modulators and terminators.

## 4. Discussions

As aforementioned, in a reverse process of the microwave CPA, the ultrathin conductive film actually behaves as a non-resonant sheet antenna that radiates the plane wave towards two sides of free space. In order to induce the oscillation of the free carriers inside the films for radiation, suppose a RF voltage bias is imposed in the film plane. According to the transmission line theory, we readily conceive an equivalent circuit which consists of a driven source with internal resistance $R_s$ accounting for the film, and two external loads $Z_0$ in parallel connection accounting for two sides of free space (the corresponding load becoming $Z_0/n$, when one half space has the refractive index $n$). It is easily to prove that the condition of the maximal power gained by the loads is the internal impedance is equal to the total external load, that is $R_s=Z_0/2$. Therefore, the CPA condition is related closely to the radiating performance of the ultrathin conductive sheet. This simple relation may be of great significance in optimizing the emission power of the radio/microwave current sheet antennas or even THz photoconductive antennas [46-48].

In technology, the thickness of the absorbing layer can be reduced further into nanometer scale while maintaining the sheet resistance. For example, a type of metallic mesh structures with thickness being in the range of several tens nanometer or a ~5nm thick smooth Au film can be used to design the 50% microwave absorber under single beam illumination [49]. Such nanofilms will give rise to the broadband perfect absorption at an even thinner dissipating scale, $\sim\lambda/10^6$, for coherent illumination.

## 5. Conclusion

In conclusion, we have experimentally demonstrated the broadband perfect absorption of the ultrathin conductive film illuminated coherently by two microwave beams. Simply using a layer of conductive coating of $R_s=Z_0/2$ and not involving resonance effect, the 100% absorption exhibits the frequency-irrelevant feature, occurs in an extremely subwavelength thickness, $\sim\lambda/10000$, and breaks the 50% limit without extra engineering of structure or substrate. The structural simplicity and the super performance cannot be achieved by the traditional approaches for EM absorbers. In addition, by adjusting the relative phase of the two microwave beams, the absorption can modulate from 0 to 1. Furthermore, no matter what value $R_s$ is, the CPA always takes place at the oblique incident cases, indicated by the zero reflection condition $R_s = (Z_0/2)\cos\theta_0$ for TM illumination or $R_s = (Z_0/2)/\cos\theta_0$ for TE illumination. Given the typical sheet resistance, the angle tolerance for the >90% absorbance is as large as 60 degree. We have also analyzed the reverse process of the microwave CPA and identified the same condition for the maximal power radiation. Our work might enable novel ultra-broadband, deep subwavelength applications in microwave absorbers and antenna engineering and bridge two important technological fields, EM absorbent materials and transparent conductors.

## Acknowledgments:

This work was supported by the National Natural Science Foundation of China (Grant Nos. 11104198, 11304215, 11104196, 11374224), the Natural Science Foundation of Jiangsu Province (Grant Nos. BK20130281, BK2011277) and a Project Funded by the Priority Academic Program Development (PAPD) of Jiangsu Higher Education Institutions. C. W. acknowledges the support from National Research Foundation for the Doctoral Program of Higher Education of China (20103201110015).

---

*Correspondence and requests for materials should be addressed to B.H. (email: houbo@suda.edu.cn)*

## References:

[1]. Jackson, J. D. *Classical Electrodynamics* (3rd ed., Wiley, 1998).

[2]. Nimtz, G. & Panten, U. Broad band electromagnetic wave absorbers designed with nano-metal films. *Ann. Phys.* **19**, 53-59 (2010).

[3]. Hilsum, C. Infrared absorption of thin metal films. *J. Opt. Soc. Am.* **44**, 188 (1954).

[4]. Hansen, R. C. & Pawlewicz, W. T. Effective conductivity and microwave reflectivity of thin metallic films. *IEEE Transactions on Microwave Theory and Techniques* **30**, 2064 (1982).

[5]. Bosman, H., Lau, Y. Y. & Gilgenbach, R. M. Microwave absorption on a thin film. *Appl. Phys. Lett.* **82**, 1353 (2003).

[6]. Staelin, D. H., Morgenthaler, A. W. & Kong, J. A. *Electromagnetic Waves*,

(Prentice-Hall, New Jersey, 1994), Sec. 4.5.

[7]. Munk, B. A. *Frequency Selective Surfaces: Theory and Design* (Wiley, New York, 2000).

[8]. Watts, C. M., Liu, X. & Padilla, W. J. Metamaterial electromagnetic wave absorbers. *Adv. Mater.* **24**, OP98–OP120 (2012).

[9]. Kats, M. A. *et al.* Ultra-thin perfect absorber employing a tunable phase change material. *Appl. Phys. Lett.* **101**, 221101 (2012).

[10]. Kats, M. A., Blanchard, R., Genevet, P. & Capasso, F. Nanometre optical coatings based on strong interference effects in highly absorbing media. *Nat. Mater.* **12**, 20–24 (2012).

[11]. Qin, F. & Brosseau, C. A review and analysis of microwave absorption in polymer composites filled with carbonaceous particles. *J. Appl. Phys.* **111**, 061301 (2012).

[12]. Chen, Z. P., Xu, C., Ma, C. Q., Ren, W. C. & Cheng, H. M. Lightweight and flexible graphene foam composites for high-performance electromagnetic interference shielding. *Adv. Mater.* **25**, 1296-1300 (2013).

[13]. Ding, F., Cui, Y. X., Ge, X. C., Jin, Y. & He, S. L. Ultra-broadband microwave metamaterial absorber. *Appl. Phys. Lett.* **100**, 103506 (2012).

[14]. Cui, Y. X., Fung, K. H., Xu, J., Ma, H., Jin, Y., He, S. L. & Fang, N. Ultrabroadband light absorption by a sawtooth anisotropic metamaterial slab. *Nano Lett.* **12**, 1443−1447 (2012).

[15]. Landy, N. I., Sajuyigbe, S., Mock, J. J., Smith, D. R. & Padilla, W. J. Perfect metamaterial absorber. *Phys. Rev. Lett.* **100**, 207402 (2008).

[16]. Li, H., Yuan, L. H., Zhou, B., Shen, X. P., Cheng, Q. & Cui, T. J. Ultrathin multiband gigahertz metamaterial absorbers. *J. Appl. Phys.* **110**, 014909 (2011).

[17]. Ye, D. X., Wang, Z. Y., Xu, K. W., Li, H., Huangfu, J. T., Wang, Z. & Ran, L. X. Ultrawideband dispersion control of a metamaterial surface for perfectly-matched-layer-like absorption. *Phys. Rev. Lett.* **111**, 187402 (2013).

[18]. Hao, J. M., Wang, J., Liu, X. L., Padilla, W. J., Zhou, L. & Qiu, M. High performance optical absorber based on a plasmonic metamateiral. *Appl. Phys.*

*Lett.* **96**, 251104 (2010).

[19]. Liu, Z. Y. *et al.* Locally resonant sonic materials. *Science* **289**, 1734-1736 (2000).

[20]. Mei, J., Ma, G. C., Yang, M., Yang, Z. Y., Wen, W. J. & Sheng, P. Dark acoustic metamaterials as super absorbers for low-frequency sound. *Nat. Commun.* **3**: 756 doi: 10.1038/ncomms1758 (2012).

[21]. Engheta, N. Thin absorbing screens using metamaterial surfaces. *IEEE AP-S International Symposium*, San Antonio, Texas, June 16-21, 2002.

[22]. Tischler, J. R., Bradley, M. S. & Bulovic, V. Critically coupled resonators in vertical geometry using a planar mirror and a 5 nm thick absorbing film. *Opt. Lett.* **31**, 2045–2047 (2006).

[23]. Dutta Gupta, S. Strong interaction mediated critical coupling at two distinct frequencies. *Opt. Lett.* **32**, 1483–1485 (2007).

[24]. Deb, S., Dutta Gupta, S., Banerji, J. & Dutta Gupta S. Critical coupling at oblique incidence. *J. Opt. A: Pure Appl. Opt.* **9**, 555–559 (2007).

[25]. Chong, Y. D., Ge, L., Cao, H. & Stone, A. D. Coherent perfect absorbers: Time-reversed lasers. *Phys. Rev. Lett.* **105**, 053901 (2010).

[26]. Wan, W. J. *et al*. Time-reversed lasing and interferometric control of absorption. *Science* **331**, 889-892 (2011).

[27]. Noh, H., Chong. Y. D., Stone, A. D. & Cao, H. Perfect coupling of light to surface plasmons by coherent absorption. *Phys. Rev. Lett.* **108**, 186805 (2012).

[28]. Noh, H., Popoff, S. M. & Cao, H. Broadband subwavelength focusing of light using a passive sink. *Opt. Express* **21**, 17435 (2013).

[29]. Dutta Gupta, S., Martin, O. J. F., Dutta Gupta, S. & Agarwal, G. S. Controllable coherent perfect absorption in a composite film. *Opt. Express* **20**, 1330-1336 (2012).

[30]. Pu, M. B. *et al*. Ultrathin broadband nearly perfect absorber with symmetrical coherent illumination. *Opt. Express* **20**, 2246-2254 (2012).

[31]. Zhang, J. F., Macdonald, K. F. & Zheludev, N. I. Controlling light-with-light without nonlinearity. *Light: Science & Applications* **1**, e18 (2012).

[32]. Kang, M. *et al*. Polarization-independent coherent perfect absorption by a dipole-like metasurface. *Opt. Lett.* **38**, 3086 (2013).

[33]. Ramakrishnan, G. *et al*. Enhanced terahertz emission by coherent optical absorption in ultrathin semiconductor films on metals. *Opt. Express* **21**, 16784 (2013).

[34]. Pirruccio, G., Moreno L. M., Lozano, G. & Rivas, J. G. Coherent and broadband enhanced optical absorption in grapheme. *ACS Nano* **7**, 4810-4817 (2013).

[35]. Chopraa, K. L., Majora, S. & Pandya, D. K. Transparent conductors—A status review. *Thin Solid Films* **102**, 1–46 (1983).

[36]. Granqvist, C. G. Transparent conductors as solar energy materials: A panoramic review. *Solar Energy Materials and Solar Cells* **91**, 1529–1598 (2007).

[37]. Gordon, R. G. Criteria for choosing transparent conductors. *MRS Bulletin* **25**, 52 (2000).

[38]. Hecht, D. S., Hu, L. B. & Irvin, G. Emerging transparent electrodes based on thin films of carbon nanotubes, graphene, and metallic nanostructures. *Adv. Mater.* **23**, 1482-1513 (2011).

[39]. Wassei, J. K. & Kaner, R. B. Graphene, a promising transparent conductor. *Materials Today* **13**, 52 (2010).

[40]. Catrysse, P. B. & Fan, S. H. Nanopatterned metallic films for use as transparent conductive electrodes in optoelectronic devices. *Nano Lett.* **10**, 2944-2949 (2010).

[41]. Zeng, X. Y., Zhang, Q. K., Yu, R. M. & Lu, C. Z. A new transparent conductor: silver nanowire film buried at the surface of a transparent polymer. *Adv. Mater.* **22**, 4484-4488 (2010).

[42]. Kuang, P. *et al*. A new architecture for transparent electrodes: Relieving the trade-off between electrical conductivity and optical transmittance. *Adv. Mater.* **23**, 2469-2473 (2011).

[43]. Wu, H. *et al*. A transparent electrode based on a metal nanotrough network. *Nature Nanotech*. **8**, 421-425 (2013).

[44]. Gordon, J. P., Zeiger, H. J. & Townes, C. H. The maser—new type of microwave

amplifier, frequency standard, and spectrometer. *Phys. Rev.* **99**, 1264-1274 (1955).

[45]. Pozar, D. M. *Microwave Engineering* (2nd ed., John Wiley & Sons, 1998).

[46]. Zhang, X. C. & Xu, J. *Introduction to THz Wave Photonics* (Springer, New York, 2010).

[47]. Lee, Y. S. *Principles of Terahertz Science and Technology* (Springer, New York, 2009).

[48]. In fact, the THz photoconductive antenna may be regarded as the radiating current sheet, where the photo-carriers generated in the optically absorbing layer of the semiconductors upon illuminated by the femtosecond laser pulse and driven by a voltage bias emit THz waves towards two sides of space.

[49]. Li, S. *et al*. Microwave absorptions of ultrathin conductive films and designs of frequency-independent ultrathin absorbers. *AIP Adv.* **4**, 017130 (2014).

# Supplementary Materials

## A. The experimental samples

The conductive coatings are widely used in transparent conductors which have far broad applications in rapidly grown optoelectronic and consumer electronic devices, for instance organic solar cells, flat-panel displays, touch screens and so on. Technically, the conductive coating with typical thickness from tens of nanometers to several micrometers can be fabricated on various substrates with its electrical property controlled, and is usually extremely thin, compared to the relevant wavelength in microwave domain. Here we utilize the customized transparent conducting films from the commercial supplier (SVG Optronics) to demonstrate the CPA phenomenon. The samples have the 0.2mm thick flexible plastic substrate and the 2.6um thick conducting layer which consists of a honeycomb lattice of conducting wires (see Supplementary Figure S1). Their sheet resistances are measured being 180Ω, 221Ω, and 442Ω by the four probe instrument.

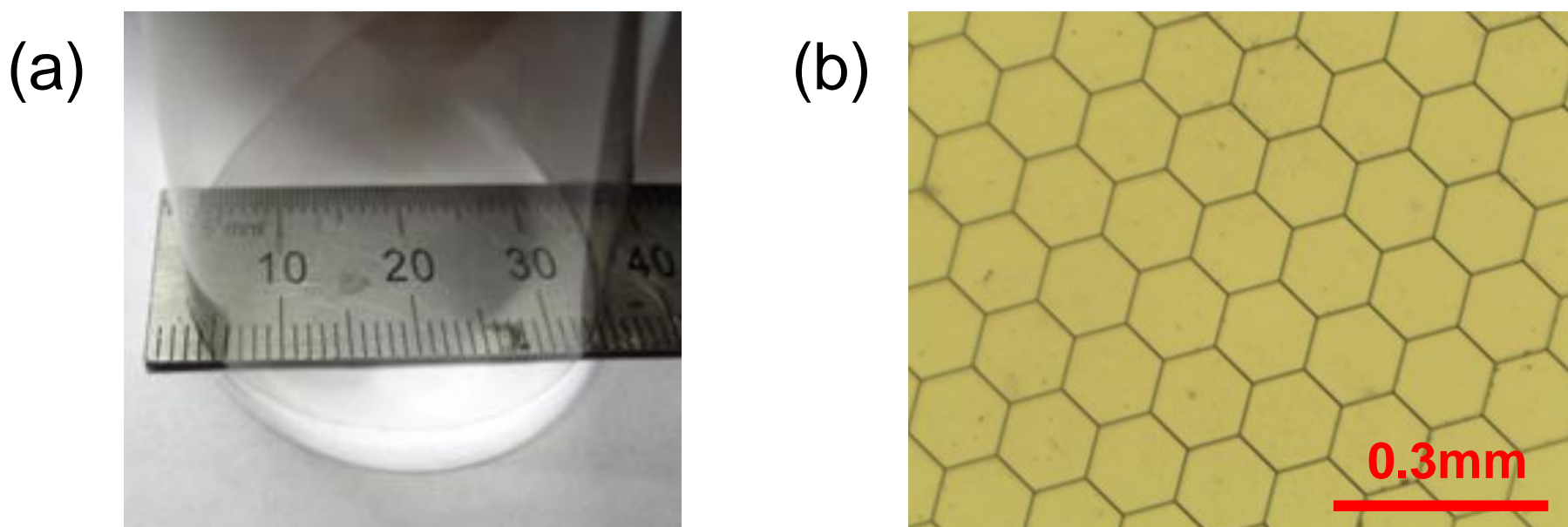

Supplementary Figure S1| The experimental sample. (a) The flexible transparent conductor sample has the 0.2mm thick plastic substrate and the 2.6um thick conducting layer. (b) The microscopic photo of the conducting layer which consists of a honeycomb lattice of conducting wires.

## B. The microwave measurements

In the measurement, we employed a microwave network analyzer (Agilent N5230C) and connected two identical rectangular horn antennas (ETS-Lindgren’s Model 3115) with port 1 and/or 2 of the network analyzer. The S-parameters, S11 and S21, are calibrated and normalized to obtain the experimental spectra. In the CPA experiments, the two horn antennas are first connected to a microwave power divider via two coaxial cables of the same length. Then the power divider is connected with port 1 of the network analyzer. The power divider behaves like the beam splitter in optics and at the same time works as the power combiner to the returned signals. The sample is inserted in the proper position with the same distance to the antennas. Before measuring the sample, we placed an 80mm thick commercial broadband absorber and calibrated S11 equal zero by subtracting. After this calibration, when an aluminum plate was mounted (PEC case) and there was nothing (OPEN case), their S11 will have the same amplitudes and the 180$^{o}$ phase difference, which can be verified by the measuring data (see Supplementary Figure S2). For the sample measurement, its S11 is normalized with respect to S11(OPEN), and is recorded as reflection amplitude. In the waveguide CPA experiment, the horn antennas were replaced by two identical coax-to-waveguide adapters (X-band WR90), and the broadband absorber was substituted with a impedance-matching waveguide termination (see Supplementary Figure S3(a)). After the calibration, the PEC and OPEN cases show the expected amplitude and phase spectra (see Supplementary Figure S3(b)).

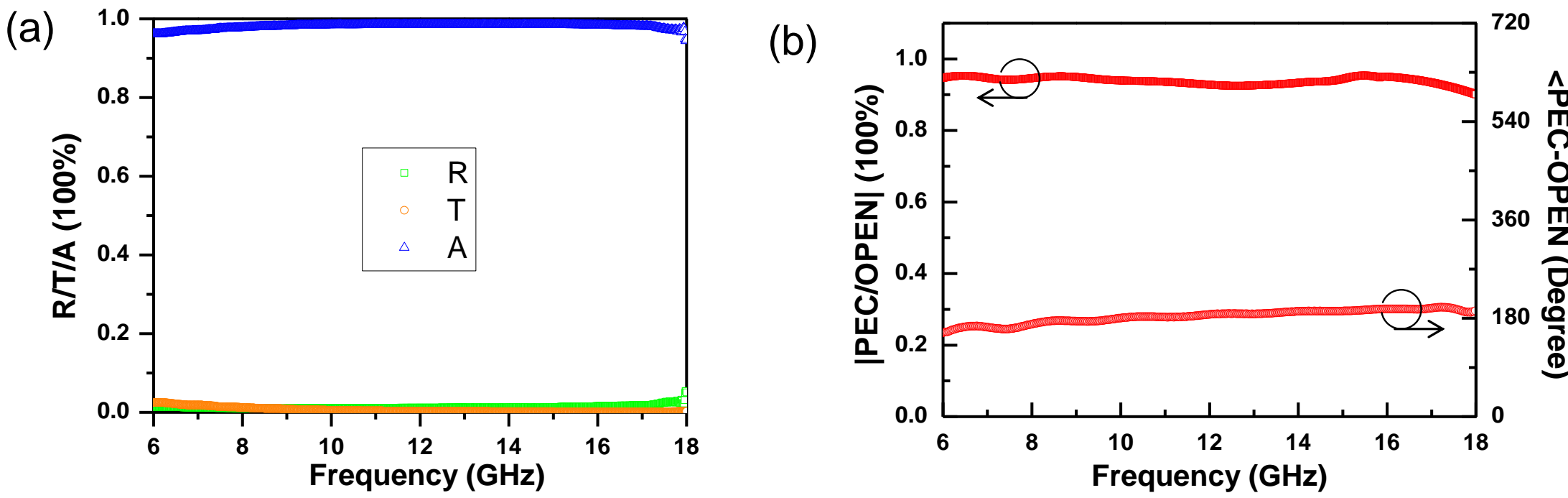

Supplementary Figure S2| The calibration for the free space measurement. (a) The reflectance, transmittance, and absorbance of the commercial broadband microwave absorber of thickness 80mm under normal incidence. (b) Left: the amplitude ratio of S11(PEC) and S11(OPEN) in the CPA measuring setup; Right: the phase difference of S11(PEC) and S11(OPEN) in the CPA measuring setup. The amplitude ratio is seen to be ~10% less than 1, which is due to the residuary reflection of the commercial absorber. The relative phase is seen to be 180 degree.

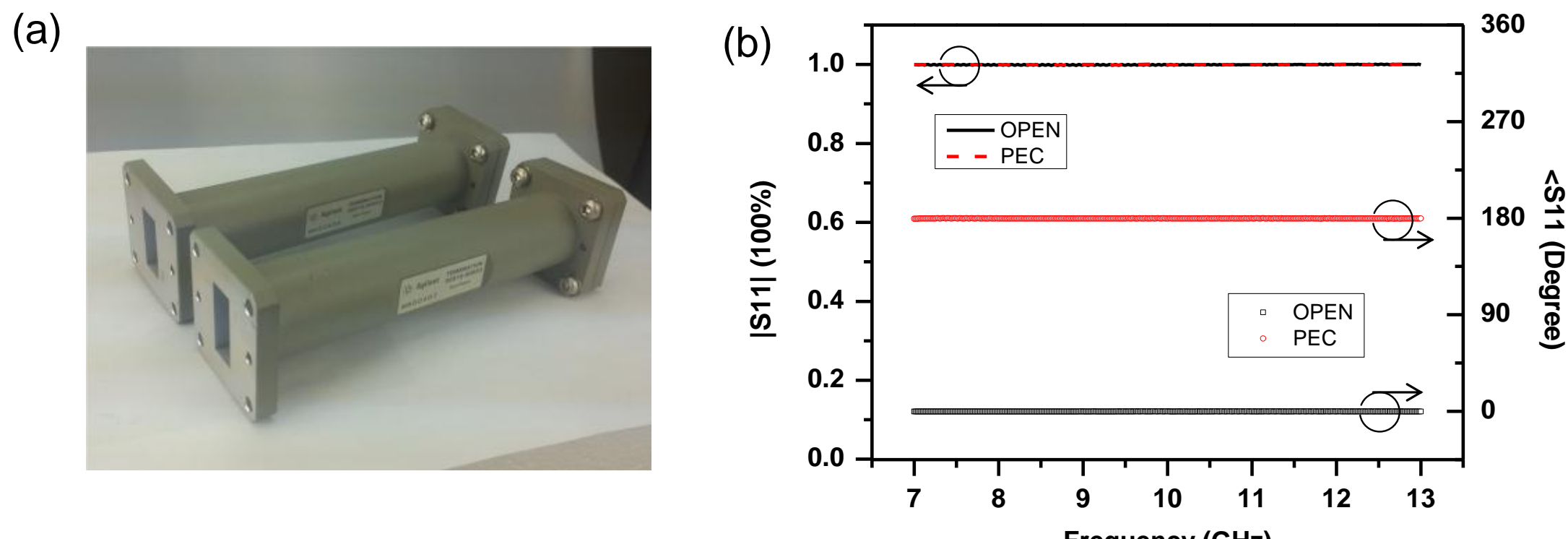

Supplementary Figure S3| The calibration for the waveguide measurement. (a) The photo of the X-band waveguide terminations; (b) Left: the amplitudes of S11(PEC) and S11(OPEN) in the waveguide measuring setup; Right: their phases.

## C. The formulas for calculating the theoretical curves

The formulas for calculating the theoretical curves in the main text come from the classical EM theory and are listed in the following where Eqs. S2 give the theoretical results in Fig. 2(b), Eqs. S3a and S3c give the theoretical maps in Figs. 3(b) and 3(c), respectively, and Eq. S4b gives the theoretical curve in Fig. 4. In these calculations, the effect of the 0.2mm plastic substrate is neglected.

(1) The absorption of the ultrathin conductive film in free space under single beam incidence and the use of the impedance language.

From the classical EM theory or Ref. [49], the transmission and reflection coefficients of the ultrathin conductive layer in free space are expressed as standard formulas in terms of permittivity or refractive index. With the impedance language, these formulas can be simplified to be:

$$t^{TE} \approx \left[1+\frac{Z_0/(2\cos\theta_0)}{R_s}\right]^{-1} \tag{S1a}$$

$$r^{TE} \approx t^{TE}\left[-\frac{Z_0/(2\cos\theta_0)}{R_s}\right] \tag{S1b}$$

$$t^{TM} \approx \left[1+\frac{Z_0\cos\theta_0/2}{R_s}\right]^{-1} \tag{S1c}$$

$$r^{TM} \approx t^{TM}\left[\frac{Z_0\cos\theta_0/2}{R_s}\right] \tag{S1d}$$

under both approximation conditions, $h<<\delta$ and $\sigma/(\omega\varepsilon_0)>>1$, and the definition $R_s=1/\sigma h$, also seeing Refs. [2-6, 49]. The two approximations are understood as ultrathin condition and low frequency condition, respectively. Here, the use of the impedance language makes the frequency-irrelevant property immediately obvious. Under normal incidence ($\theta_0=0$), $r^{TE}=-r^{TM}=r$ and $t^{TE}=t^{TM}=t$. The absorbance is calculated via the relation $A=1-R-T$, where the reflectance $R=|r^{TE(TM)}|^2$ and the transmittance $T=|t^{TE(TM)}|^2$.

(2) The coherent absorption of the ultrathin conductive film in free space at normal incidence.

As mentioned in the main text, all scattered signals, including $r1$, $t1$, $r2$, and $t2$, are considered as reflection. From the splitting and combining of the signals inside the -3dB power divider (the broadband Wilkinson type), the reflectance is calculated as:

$$R=|0.5*(r1+t2)+0.5*(r2+t1)|^2=|r+t|^2 \\ =\frac{|1-\frac{Z_0}{2R_s}|^2}{|1+\frac{Z_0}{2R_s}|^2}, \tag{S2a}$$

where $r1=r2=r$ and $t1=t2=t$ are applied and $L=0$ is assumed. The absorbance is calculated as:

$$A=1-R=1-\frac{|1-\frac{Z_0}{2R_s}|^2}{|1+\frac{Z_0}{2R_s}|^2}. \tag{S2b}$$

If the distances of the sample to the two horn antennas are adjusted so that $L$ is nonzero, then

$$R=|0.5e^{ik_0L}(1+r+re^{ik_0L})+0.5(r+e^{ik_0L}+re^{ik_0L})|^2; \tag{S2c}$$

$$A=1-R=1-|0.5e^{ik_0L}(1+r+re^{ik_0L})+0.5(r+e^{ik_0L}+re^{ik_0L})|^2, \tag{S2d}$$

where $1+r=t$ is applied.

(3) The coherent absorption of the ultrathin conductive film in free space at oblique incidence.

For TM mode:

$$R=\frac{|1-\dfrac{0.5Z_0\cos\theta_0}{R_s}|^2}{|1+\dfrac{0.5Z_0\cos\theta_0}{R_s}|^2};\tag{S3a}$$

$$A=1-R=1-\frac{|1-\dfrac{0.5Z_0\cos\theta_0}{R_s}|^2}{|1+\dfrac{0.5Z_0\cos\theta_0}{R_s}|^2}.\tag{S3b}$$

For TE mode:

$$R=\frac{|1-\dfrac{0.5Z_0/\cos\theta_0}{R_s}|^2}{|1+\dfrac{0.5Z_0/\cos\theta_0}{R_s}|^2};\tag{S3c}$$

$$A=1-R=1-\frac{|1-\dfrac{0.5Z_0/\cos\theta_0}{R_s}|^2}{|1+\dfrac{0.5Z_0/\cos\theta_0}{R_s}|^2}.\tag{S3d}$$

(4) The coherent absorption of the ultrathin conductive film in the waveguide with phase delay.

$$\begin{aligned}R&=|0.5e^{i\beta L}(t^{TE}+r^{TE}e^{i\beta L})+0.5(r^{TE}+t^{TE}e^{i\beta L})|^2\\&=\frac{|1-\dfrac{0.5Z_0/\cos\theta_0}{R_s}\cos(\beta L)|^2}{|1+\dfrac{0.5Z_0/\cos\theta_0}{R_s}|^2}\end{aligned}\tag{S4a}$$

$$A=1-R=1-\frac{|1-\dfrac{0.5Z_0/\cos\theta_0}{R_s}\cos(\beta L)|^2}{|1+\dfrac{0.5Z_0/\cos\theta_0}{R_s}|^2}\tag{S4b}$$

where $t^{TE}$ (Eq.S1a) and $r^{TE}$ (Eq.S1b) are applied with $\cos\theta_0=\beta/k_0$, $\beta=\sqrt{k_0^2-(\pi/a)^2}$ (the side length of the waveguide cross-section $a$=22.86mm) denotes the propagation constant of TE10 mode, and $L$=140mm. From Eq. S4a, we obtain the $R$=1 condition by the numerator equating the denominator, i.e. $\cos(\beta L)=-1$, and the $R$=0 condition by the numerator equating zero, i.e. $0.5Z_0\cos(\beta L)/(R_s\cos\theta_0)=1$.

**D. The EM parameters of the experimental film sample**

The EM parameters, such as permittivity or refractive index, of the ultrathin conductive film can be retrieved from the transmission/reflection coefficients. Specific to the transparent conductor sample, we consider a retrieving problem of a two-layered film which consists of a conductive layer and a dielectric substrate.

In the case of normal incidence of the EM wave in free space, the transmission coefficient of the two-layered film can be calculated to be

$$t = \frac{-2}{i(Z_1' + 1/Z_1')\sin\delta_1\cos\delta_2 + i(Z_2' + 1/Z_2')\cos\delta_1\sin\delta_2 + (Z_1'/Z_2' + Z_2'/Z_1')\sin\delta_1\sin\delta_2 - 2\cos\delta_1\cos\delta_2},$$

where $\delta_1 = n_1 k_0 h_1$, $\delta_2 = n_2 k_0 h_2$, $Z_1' = \frac{Z_1}{Z_0}$, $Z_2' = \frac{Z_2}{Z_0}$, $k_0$ is the vacuum wave vector, $n_1$ ($n_2$) is the refractive index of the conductive layer (substrate), $h_1$ ($h_2$) is the thickness of the conductive layer (substrate), and $Z_0$, $Z_1$ and $Z_2$ are the wave impedance in free space, the conductive layer and the substrate, respectively. For our sample, the relative permittivity and permeability of the dielectric layer, $\varepsilon_{r2} = 2.0$ and $\mu_{r2} = 1$, are readily obtained from measuring the bare plastic substrate without the conductive layer. Since the ultrathin conductive layer is nonmagnetic, its relative permeability $\mu_{r1} = 1$, which implies $\varepsilon_{r1}$ is the only unknown. Given $h_1$=2.6um, $h_2$=0.2mm, and the transmission $t$, we can solve the above equation to retrieve the relative permittivity $\varepsilon_{r1}$ of the conductive layer without concerning the reflection coefficient.

Taking the film sample of $R_s$=180Ω as an example, we have plotted its S21 spectra in Supplementary Figure S4(a), and understand that S21 is the transmission coefficient except the opposite sign convention of the phase. The relative permittivity $\varepsilon_{r1}$ of the conductive layer is retrieved and shown in Supplementary Figure S4(b). It is noticed that the imaginary part of $\varepsilon_{r1}$ dominates over the real part in the experimental frequency regime. Consequently, the real part and the imaginary part of the refractive index $n_1 = \sqrt{\varepsilon_{r1}}$ are almost equal, as seen in Supplementary Figure S4(c). Then, the conductivity is obtained via the relation, $\varepsilon_{r1} \approx i\sigma/\omega\varepsilon_0$, in Supplementary Figure S4(d). Also, we have performed the COMSOL Multiphysics simulation which agrees very well with the experimental results. Particularly, the experimental value (symbols in Supplementary Figure S4(d)) of the conductivity is nondispersive, matching the prescribed constant (dash line in Supplementary Figure S4(d)) in the simulation.

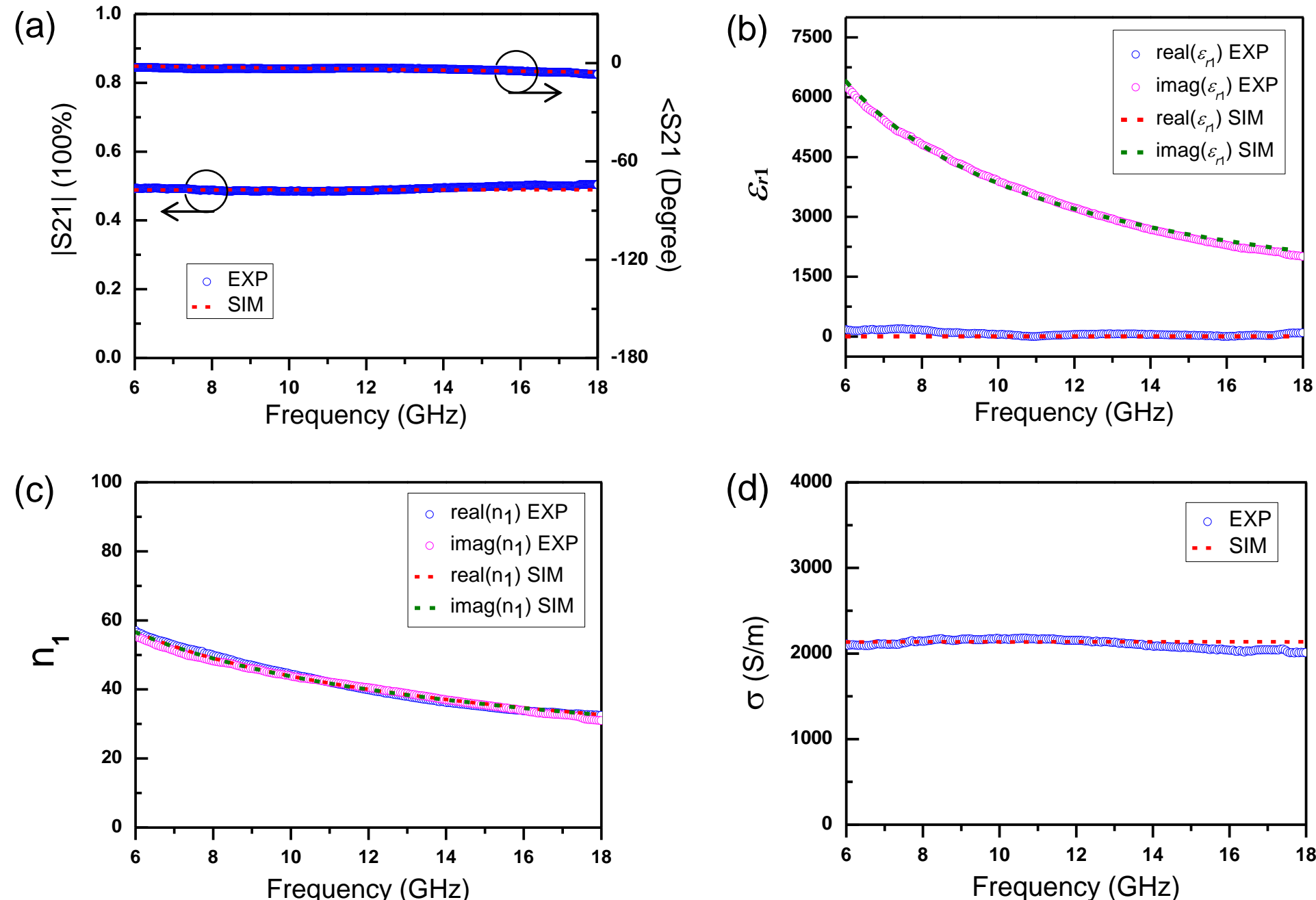

Supplementary Figure S4| The retrieved electromagnetic parameters of the film sample $R_s$=180Ω. (a) The S21 amplitude and phase in experiment (symbols) and simulation (dash line). (b) The real and imaginary part of relative permittivity $\varepsilon_{r1}$ retrieved from the experimental and simulated S21 parameter. (c) The real and imaginary part of refractive index $n_1$. (d) The conductivity $\sigma$.

## E. The discussion about the physical origin of the twin absorption peaks

In order to understand clearly the origin of the twin peaks around the condition $\Delta\varphi=\beta L=m\pi$ ($m$ being the even number), we subtotal $r1$ and $t2$ ($r2$ and $t1$) to represent the signal scattered toward the left (right) side of the film, and obtain $0.5e^{i\beta L}(t^{TE} + r^{TE}e^{i\beta L})$ for the left signal and $0.5(r^{TE} + t^{TE}e^{i\beta L})$ for the right signal. As shown in Supplementary Figure S5(b), their amplitudes are equal to each other and exhibit the modulation lineshape caused by the interference between $r1$ and $t2$ ($r2$ and $t1$) and controlled by the delay phase. The amplitude maxima correspond to the condition $\Delta\varphi=\beta L=m\pi$ ($m$ being the odd number), and the dips appear when $\Delta\varphi=\beta L=m\pi$ ($m$ being the even number). Exactly at these delay phase values, the two ways of signals return to the power divider with their phase difference being 0. However, around the nonzero dips ($\beta L=2\pi$, $4\pi$, $6\pi$, and $8\pi$ within our measuring frequency band) due to the imperfect cancellation between $r1$ and $t2$ ($r2$ and $t1$), the relative phase relationship of the two ways of signals may become totally out of phase ($180^{o}$) at two frequencies located on both sides of these dips. While they meet to combine inside the power divider, they will cancel each other because of the same amplitude and the $180^{o}$ phase difference. Therefore, we observe the zero reflection at the two frequencies associated with every nonzero dips, which leads to the twin absorption peaks.

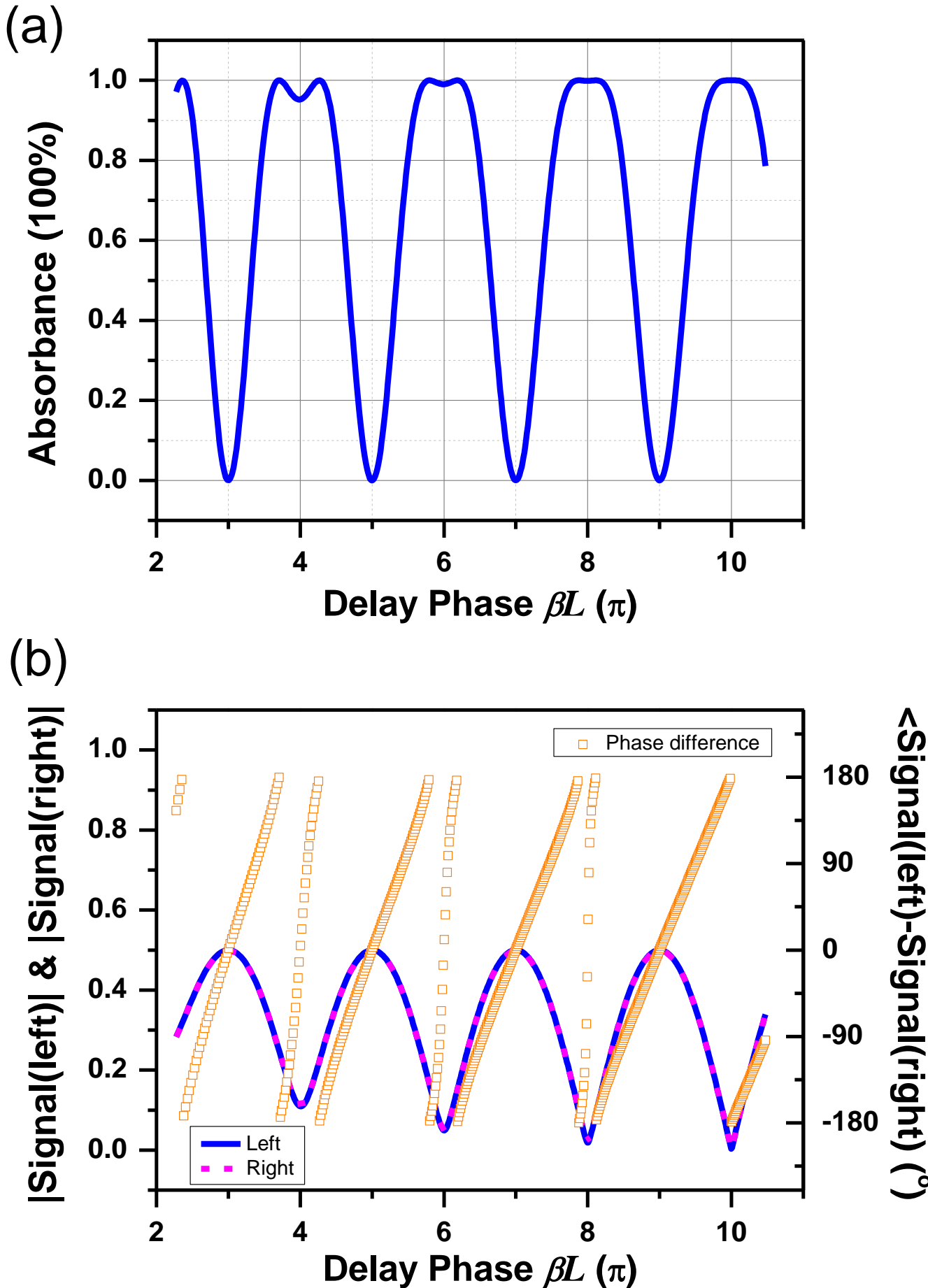

Supplementary Figure S5| (a) The coherent absorbance in the waveguide with the $L$=140mm long delay line, calculated according to Eq. S4b and plotted versus the delay phase. (b) Left: the amplitudes of the signals scattered toward the left and right side of the film, respectively. Right: the phase difference between the two ways of the signals when they return to the power divider. The sheet resistance takes 221Ω.